\documentclass[aps,pra,twocolumn,
nofootinbib,showpacs]{revtex4}

\usepackage{amsmath, latexsym, amssymb, amscd,amsthm}
\usepackage{graphicx}
\usepackage{pictex}

\newtheorem{thm}{\bf Theorem}[]

\begin{document}

\title{
Fisher information and asymptotic normality in system identification for  
quantum Markov chains}

\author{M\u{a}d\u{a}lin Gu\c{t}\u{a}}
\affiliation{University of Nottingham,\\School of Mathematical Sciences, \\
University Park, Nottingham NG7 2RD, UK}

\begin{abstract}

This paper deals with the problem of estimating the coupling constant $\theta$ 
of a mixing quantum Markov chain. For a repeated measurement on the chain's output we show that the outcomes' time average has an asymptotically normal (Gaussian) distribution, and we give the explicit expressions of its mean and variance. In particular we obtain a simple estimator of $\theta$ whose {\it classical} Fisher information can be optimized over different choices of measured observables. We then show that the quantum state of the output together with the system, is itself asymptotically Gaussian and compute its {\it quantum} Fisher information which sets an absolute bound to the estimation error. The classical and quantum Fisher informations are compared in a simple example. In the vicinity of $\theta=0$ we find that the quantum Fisher information has a quadratic rather than linear scaling in output size, and asymptotically the Fisher information is localised in the system, 
while the output is independent of the parameter.
%
%

\end{abstract}
\maketitle
\section{Introduction}

Quantum Statistics started in the 70's with the discovery that notions of `classical' statistics such as the Cram\'{e}r-Rao inequality, the Fisher information, have non-trivial quantum extensions which can be used to design optimal measurements for quantum state estimation and discrimination \cite{Holevo,Helstrom,Belavkin,Yuen&Lax}. Recently, statistical inference has become an indispensable tool in quantum engineering tasks such as state preparation \cite{Haffner,Breitenbach}
, precision metrology \cite{Giovanetti,Higgins}, quantum process tomography 
\cite{
Lobino,Brune}, state transfer and teleportation  \cite{
Polzik,telepolzik}, continuous variables tomography \cite{Vogel&Risken,Lutterbach}.

Quantum system identification (QSI) is a topic of particular importance in quantum engineering and control \cite{Khaneja&Mabuchi} where accurate knowledge of dynamical parameters is crucial. This paper addresses the QSI problem for Markov dynamics from the viewpoint of {\it asymptotic statistics},  complementing other recent investigations \cite{Mabuchi,Howard,Schirmer&Oi,Burgarth}.

\begin{figure}[h]
\begin{center}
\includegraphics[width=8cm]{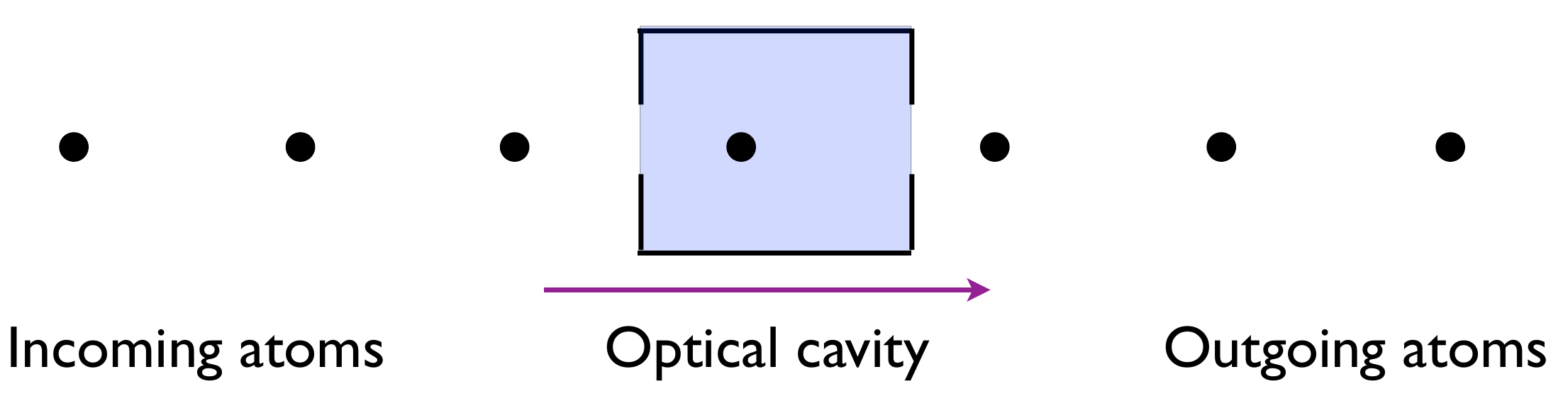}
\caption{Atom maser: identically prepared atoms pass successively through a cavity, interact with the cavity field, with the outgoing atoms carrying information on the dynamics.}
\label{fig.atom.maser}
\end{center}
\end{figure}

We illustrate the concept of a quantum Markov chain through the example of an atom maser \cite{Brune}: identically prepared d-level atoms (input) pass successively and at equal time intervals $\tau$ through a cavity, interact with the cavity field, and exit in a perturbed state  (output) which carries information about the interaction (see Figure \ref{fig.atom.maser}). Neglecting the internal dynamics of atoms and cavity, and taking the latter to be of dimension 
$k<\infty$, the evolution can be described in discrete time and consists of applying an interaction unitary $U\in M(\mathbb{C}^{d}\otimes \mathbb{C}^{k})$ for each time interval  $\tau$ when a new atom passes through the cavity. If the incoming atoms are in the pure state $\psi\in \mathbb{C}^{d}$ and the cavity is initially in some state $\varphi\in\mathbb{C}^{k}$, then at time $n$ the output plus cavity state is 
$\psi^{n}\in  \left(\mathbb{C}^{d}\right)^{\otimes n}\otimes \mathbb{C}^{k}$
\begin{equation}\label{eq.output}
\psi^{n}:= U(n) \, \left(\psi^{\otimes n}\otimes \varphi \right):= 
U^{(1)} \cdots U^{(n)} \, \left(\psi^{\otimes n}\otimes \varphi\right)
\end{equation}
where $U^{(l)}$ is the copy of $U$ acting on cavity and the atom which at time $n$ is at position $l$ on the right side of the cavity.

Consider now that the interaction depends on some unknown parameter 
$\theta$ such that $U=U_{\theta}$ and correspondingly $\psi^{n}=\psi^{n}_{\theta}$. Our identification problem is to 
estimate $\theta$, by measuring the {\it output} rather that the system (cavity) which may not be directly accessible. For simplicity we restrict ourselves to the case of estimating one parameter (the interaction strength), such that $U_{\theta}= \exp(-i\theta H)$ where $H\in M(\mathbb{C}^{d}\otimes \mathbb{C}^{k})$ is a {\it known} hamiltonian,
but the results can be extended to multiple parameters as well as continuous time \cite{Guta&Bouten}. 
The questions we want to address are: how much information about $\theta$ 
is contained in the output state $\psi_{\theta}^{n}$, 
and how can we `extract' it ?

The standard approach to such questions goes via the quantum Cram\'{e}r-Rao inequality which shows that the variance of any unbiased estimator obtained by measuring a copy of a state $\rho_{\theta}$, is lower  bounded by the inverse of the quantum Fisher information $F(\theta)^{-1}$ \cite{Helstrom,Holevo,Braunstein&Caves}.  Although this bound is generally not attainable for a single copy, it is asymptotically attainable, i.e. there exist a sequence of measurements $M_{n}$ on 
$n$ {\it identically prepared systems}, and estimators $\hat{\theta}_{n}$ such that
$$
\lim_{n\to\infty} n \mathbb{E} (\hat{\theta}_{n}-\theta)^{2} = F(\theta)^{-1}.
$$
Unlike this case where the $n$-copies Fisher information scales linearly with the number of systems, the Fisher information of the {\it correlated} states 
$\psi^{n}_{\theta}$ depends on the joint state rather than that of a single sub-system, and there is no straightforward argument to show that its rescaled version is asymptotically attainable. Moreover, in the case of multi-dimensional parameters, this approach would run into the same problems as the independent copies model,  for which it is well known that  the Cram\'{e}r-Rao bound is not attainable even asymptotically \cite{Holevo}.


For these reasons, we will pursue an {\it asymptotic} analysis based on the concept of local asymptotic normality (LAN) \cite{vanderVaart} which was recently extended to quantum statistics \cite{Guta&Kahn,Guta&Jencova,Guta&Janssens&Kahn,Guta&Kahn2} and used to solve the (asymptotically) optimal estimation problem for general  multiparametric models. In this paper we extend quantum LAN from independent to finitely correlated quantum states, and in the same time generalise results on  LAN \cite{Hopfner} and the Central Limit Theorem (CLT) for  classical Markov chains \cite{Meyn&Tweedie}.


%

In Theorem  \ref{th.qlan}  we show that, 
locally with respect to the parameter $\theta$, the output state $\psi^{n}_{\theta}$ can be approximated by a one-parameter family of coherent states, cf. \eqref{eq.model.convergence}. This result can be seen as the Markov version of asymptotic Gaussianity in coherent spin states (CSS) \cite{Radcliffe}. As a by-product, we obtain the asymptotic expression of the 
 quantum Fisher information $F= F(\theta)$ (per atom) 
 of the output state \eqref{eq.q.fisher.info}, which is equal to the `Markovian variance' of the driving hamiltonian. 
 This quantity should be understood as the limit of the rescaled quantum Fisher informations $F^{(n)}(\theta)$  of the $n$ atoms family of states $\psi^{n}_{\theta}$
 $$
 F (\theta) = \lim_{n\to\infty} \frac{F^{(n)} (\theta)}{n}.
 $$
In particular, $F(\theta)^{-1}$ sets an asymptotic lower bound on the variance $n \mathbb{E} (\hat{\theta}_{n}-\theta)^{2}$ of any sequence of unbiased estimators $\{\hat{\theta}_{n} \}$.
Using the formalism of LAN we can show that the lower bound is attainable but at the moment we do not not know the  explicit form of the optimal measurement, and we expect it to be non-separable, and possibly unfeasible with current technology.

In Theorem \ref{th.simple.measurement} we analyse the more realistic set-up of  repeated, separate measurements performed on the outgoing atoms, and show  that 
the {\it mean} of the outcomes is asymptotically Gaussian and can be used to estimate $\theta$. The corresponding {\it classical} Fisher information can be maximised over different measured observables, so that the experimenter can perform the most informative separable measurement, and compare its performance with the benchmark given by the quantum Fisher information (see Example). This generalises Wiseman's adaptive phase estimation protocol where a particular field quadrature is most informative among all quadratures \cite{Wiseman}. With the same techniques,  similar results can be obtained for means of other functionals of the measurement data such as correlations between subsequent atoms. However it remains an open problem to find the (asymptotic) classical Fisher information contained in the complete measurement data.

The next two sections introduce the key concepts underlying our results: local asymptotic normality and ergodicity. The main results are contained in Theorems 3 and 4. To illustrate these results we analyse a simple example based on a XY interaction for which we compute the quantum Fisher information and we plot the classical Fisher information for different output observables. An interesting feature of this model is the divergence of the asymptotic quantum Fisher information per atom at vanishing  interaction, which is due to a quadratic rather than linear scaling of the `usual' quantum Fisher information $F^{(n)}(\theta)$ for $\theta\approx 0$. We investigate this behaviour and find that with the scaling $\theta=u/n$, the model converges to a simple unitary rotation model on the system, with input passing into the output unperturbed. We conclude with a discussion on further extensions, open problems and connections with other topics.

\section{Local asymptotic normality in classical and quantum statistics}
\label{sec.2}

In this section we briefly review some asymptotic statistics techniques, show how they extend to quantum statistics, and explain why this is useful. The aim is to introduce the concept of local asymptotic normality, which will be encountered in the main results, Theorems \ref{th.simple.measurement} and \ref{th.qlan}.


\subsection{Asymptotic estimation in classical statistics}

A typical problem in statistics is the following: estimate an unknown parameter $\theta= (\theta_{1},\dots,\theta_{p}) \in\mathbb{R}^{p}$, given the random variables ${\bf X}_{1},\dots ,{\bf X}_{n}$ which are independent and identically distributed (i.i.d.) and have probability distribution $\mathbb{P}_{\theta}$ depending `smoothly' on $\theta$. 
If $\hat{\theta}_{n}:=\hat{\theta}_{n}({\bf X}_{1},\dots ,{\bf X}_{n})$ is a unbiased estimator, that is $\mathbb{E}(\hat{\theta}_{n})= \theta$, then the Cram\'{e}r-Rao (C-R)
inequality provides the following lower bound to its (rescaled) 
covariance matrix
\begin{equation}\label{eq.classical.CR}
n\mathbb{E}\left[( \hat{\theta}_{n} - \theta)^{T}( \hat{\theta}_{n} - \theta)\right] \geq I(\theta)^{-1},
\end{equation}
where $I(\theta)$ is a $p\times p$ positive definite real matrix called the 
Fisher information matrix at $\theta$ and quantifies the amount of `statistical information' about $\theta$ contained in a single sample from the distribution 
$\mathbb{P}_{\theta}$. If $p_{\theta}(x)=d\mathbb{P}_{\theta}/d\mu$ denotes the  density of $\mathbb{P}_{\theta}$ with respect to some reference measure 
$\mu$, then the matrix elements of $I(\theta)$ are given by 
$$
I(\theta)_{ij}:=
\int p_{\theta}(x) \frac{\partial \log p_{\theta}(x)}{\partial \theta_{i} }  
\frac{\partial \log p_{\theta}(x)}{\partial \theta_{j} } \mu(dx) =
\mathbb{E}_{\theta}(\ell_{\theta,i} \ell_{\theta,j})
$$ 
and depends only on the local behavior of the statistical model 
$\{\mathbb{P}_{\theta} :\theta \in \mathbb{R}^{p}\}$ around the point $\theta$. 
To give a simple example, if $X_{i}\in \{0,1\}$ are independent coin tosses 
with $\mathbb{P}_{\theta}[X_{i}=1]=\theta$, then the mean 
$$
\hat{\theta}_{n} =\frac{1}{n}\sum_{i=1}^{n} X_{i}
$$
is an unbiased estimator of $\theta$ whose distribution is asymptotically 
normal according to the Central Limit Theorem  
$$
\sqrt{n}(\hat{\theta}_{n}-\theta) \overset{\mathcal{L}}{\longrightarrow} N(0, \theta(1-\theta)).
$$
where $N(m,V)$ is the normal distribution with mean $m$ and variance $V$. 
A simple calculation shows that in this case $I(\theta)^{-1}= \theta(1-\theta)$ so that $\hat{\theta}_{n}$ achieves the C-R lower bound. 
Interestingly, the Fisher information diverges at the boundary of the interval $[0,1]$, but note that the bound is meaningful only for points in the interior of 
the parameter space. A similar situation will occur later in an example of a quantum Markov chain.

While in general there might not exist any unbiased estimators achieving the 
C-R bound for a given $n$, the theory says that `good' estimators $\hat{\theta}_{n}({\bf X}_{1},\dots ,{\bf X}_{n})$ (e.g. maximum likelihood under certain conditions) are {\it asymptotically} normal with
\begin{equation}\label{eq.mle}
\sqrt{n} (\hat{\theta}_{n}-\theta) \overset{\mathcal{L}}{\longrightarrow} N(0, I(\theta)^{-1})
\end{equation}
such that the C-R bound is asymptotically achieved \cite{Lehmann&Casella}.

Le Cam went a step further and discovered a more fundamental phenomenon called {\it local asymptotic normality} (LAN), which roughly means that the underlying statistical model $\{\mathbb{P}_{\theta} :\theta \in\mathbb{R}^{p}\}$ can be  linearised in the neighbourhood of any fixed parameter, and approximated by a simple Gaussian model with fixed covariance and unknown mean \cite{LeCam}. To explain this, let us first note that without loss of generality we can `localise' $\theta$, i.e. write it as
$$
\theta= \theta_{0} + u/\sqrt{n}
$$ 
where $\theta_{0}$ can be chosen to be a rough estimator based on a small sub-sample of size $\tilde{n}= n^{1-\epsilon}$ with $0<\epsilon\ll 1$, 
and $u$ is an unknown `local parameter'. By a simple concentration of measure argument \cite{Guta&Janssens&Kahn} one can show that with vanishing probability of error, the local parameter satisfies 
$\|u\|\leq n^{\epsilon}$. For all practical purposes, we can then use the more convenient local parametrisation by $u\in \mathbb{R}^{p}$ and denote the original distribution $\mathbb{P}^{n}_{\theta_{0} + u/\sqrt{n}}$ by $\mathbb{P}_{n,u}$. Now, LAN is the statement that
there exists randomisation (classical channels) $T_{n}$ and $S_{n}$ such that
\begin{eqnarray*}
&&
\lim_{n\to\infty}
\sup_{\|u\|\leq n^{\epsilon}} 
\| T_{n}(p_{n,u})  -  \mathcal{N}(u)\|_{1}
=0
\\
&&
\lim_{n\to\infty}
\sup_{\|u\|\leq n^{\epsilon}} 
\| p_{n,u}- S_{n}( \mathcal{N}(u))\|_{1}=0
\end{eqnarray*}
where $\| \cdot \|_{1}$ denotes the $L_{1}$-norm, and $p_{n,u}$ and 
$\mathcal{N}(u)$ are the probability densities of $\mathbb{P}_{n,u}$ and respectively $N(u, I(\theta_{0})^{-1})$. Operationally this means that one can use the data ${\bf X}_{1},\dots ,{\bf X}_{n}$ to simulate a normally distributed variable with density $\mathcal{N}(u)$, and viceversa, with asymptotically vanishing $L_{1}$-error, without having access to the unknown parameter $u$. This type of convergence is strong enough to imply the previous results on asymptotic normality and optimality of the maximum likelihood estimation, but can be used to make similar statements about other statistical decision problem  concerning $\theta$. We will now show that a similar phenomenon occurs in quantum statistics, and indicate how it can be used to finding asymptotically optimal state estimation protocols. 

\subsection{The Cram\'{e}r-Rao approach to state estimation}

Let us consider the problem of estimating  a one-dimensional parameter $\theta$, given $n$ {\it identical and independent} copies of a quantum system prepared in the (possibly mixed) state 
$
\rho_{\theta} \in M(\mathbb{C}^{d})
$
depending smoothly on $\theta$. Following  \cite{Holevo,Helstrom,Belavkin,Yuen&Lax,Braunstein&Caves} we analyse an {\it unbiased} estimator $\hat{\theta}_{n}$  based on the outcome of an arbitrary measurement $M$ on the joint state 
$\rho_{\theta}^{\otimes n}$. By the classical Cram\'{e}r-Rao inequality, the mean square error (MSE) of  $\hat{\theta}_{n}$ is lower bounded by the inverse (classical) Fisher information of the measurement outcome:
$$
\mathbb{E} \left[(\hat{\theta}_{n}-\theta)^{2}\right] \geq I_{M}(\theta)^{-1}.
$$
Thus, a `good' measurement is characterised by a large Fisher information, but how large can $I_{M}(\theta)$ be ? The answer is given by the notion of {\it quantum Fisher information} associated to the model 
$\{\rho_{\theta}: \theta\in \mathbb{R}\}$ which is defined as  
$$
F(\theta):= {\rm Tr}(\rho_{\theta} \mathcal{L}_{\theta}^{2}),\qquad 
$$
where the {\it symmetric logaritmic derivative} (sld) $\mathcal{L}_{\theta}$ is the quantum analogue of $\ell_{\theta}$ and is the selfadjoint solution of the equation
$$
\frac{d \rho_{\theta}}{d\theta}=
\frac{1}{2}\left( \mathcal{L}_{\theta} \rho_{\theta}+ \rho_{\theta}\mathcal{L}_{\theta}\right).
$$
As expected, for $n$ identical copies the  quantum Fisher information of the joint state $\{\rho_{\theta}^{\otimes n}: \theta\in \mathbb{R}\}$ is 
$nF(\theta)$ and the sld is given by
$$
\mathcal{L}_{\theta}(n)= \sum_{i=1}^{n} \mathcal{L}_{\theta}^{(i)},
$$
with $\mathcal{L}_{\theta}^{(i)}$ acting on the $i$'th system.

The Braunstein-Caves inequality shows that the quantum Fisher information is an upper bound to the classical one, i.e.
$
I_{M}(\theta) \leq n F(\theta)
$
so that we obtain
$$
n \mathbb{E} \left[(\hat{\theta}_{n}-\theta)^{2}\right] \geq F(\theta)^{-1}.
$$
Moreover, the constant on the right side can be achieved {\it asymptotically} by an adaptive measurement which amounts to measuring  
$\mathcal{L}_{\theta_{0}}(n)$ for some point $\theta_{0}$ which is a preliminary estimator of $\theta$ obtained by measuring a small proportion of 
the systems. In summary, the optimal estimation rate for one dimensional parameters is $n^{-1}F(\theta)^{-1}$, and can be achieved by means of 
{\it separate} measurements of the sld's $\mathcal{L}_{\theta}^{(i)}$. 

Let us turn now to the case where $\rho_{\theta}$ depends on a multidimensional parameter $\theta= (\theta_{1},\dots ,\theta_{p})\in \mathbb{R}^{p}$. The quantum Fisher information {\it matrix} can be defined along similar lines and all previous inequalities hold as matrix inequalities, in particular the covariance matrix of an unbiased estimator $\hat{\theta}_{n}$ satisfies
\begin{equation}\label{eq.cramer.rao}
n \mathbb{E} 
\left[ 
(\hat{\theta}_{n} - \theta)^{T} (\hat{\theta}_{n} - \theta)
\right] \geq F(\theta)^{-1}.
\end{equation}
However, unlike the classical case, and 
the one-dimensional quantum case, the right side is in general not achievable, even asymptotically ! This purely quantum phenomenon has a simple intuitive explanation: the optimal estimation of the different coordinates 
$(\theta_{1},\dots,\theta_{p})$ requires the simultaneous measurement of  generally incompatible observables, the associated sld's 
$(\mathcal{L}_{\theta,1}(n),\dots ,\mathcal{L}_{\theta,p}(n))$. 

Coming back to the original goal of estimating the parameter $\theta$, 
the above Fisher information analysis implies that the optimal measurement procedures must depend on the chosen figure of merit. Thus, one should not aim at saturating matrix inequalities such as \eqref{eq.cramer.rao} but at finding asymptotically attainable lower bounds 
for the risk (multiplied by $n$)
$$
n\mathbb{E} \left[ (\hat{\theta}_{n} - \theta) G(\theta) (\hat{\theta}_{n} - \theta)^{T}\right]= n{\rm Tr}\left( G(\theta) {\rm Var}(\hat\theta_{n})\right),
$$ 
assuming for simplicity a quadratic loss function with positive weight matrix $G(\theta)$. Taking the trace with $G(\theta)$ on both sides of \eqref{eq.cramer.rao} gives the generally non-attainable lower bound 
${\rm Tr}(G(\theta)F(\theta)^{-1})$, and other examples can be derived from different versions of the quantum Cram\'{e}r-Rao inequality such as Belavkin's right and left inequalities \cite{Belavkin}. Holevo \cite{Holevo} derived a more general bound and showed that it is achievable for families of Gaussian states with unknown displacements, but until recently it remained an open question whether this bound was asymptotically attainable for finite dimensional states. By further refining the techniques of the unbiased estimation set-up, Hayashi and Matsumoto \cite{Hayashi&Matsumoto} showed that the Holevo bound is indeed asymptotically attainable for general families of 
{\it two-dimensional} quantum states. Complementing this frequentist asymptotic analysis, Bagan and coworkers \cite{Bagan&Gill} solved the optimal qubit estimation problem {\it for any given $n$} in the Bayesian set-up with invariant priors. However, neither of these approaches was successful in solving the (asymptotically) optimal estimation problem for general (mixed states), multi-parametric models $\rho_{\theta} \in M(\mathbb{C}^{k}) $ with $k>2$.

\subsection{Local asymptotic normality for quantum states} 

 At this point, a natural question to ask is whether the phenomenon of local asymptotic normality occurs also in quantum statistics, and whether it can be used to design asymptotically optimal measurement strategies. Recall that in the classical case, the main idea was that for large $n$, the i.i.d. model could be approximated by a Gaussian model, in the sense that each can be mapped approximately into the other by means of classical channels. Building on earlier work by Hayashi \cite{Hayashi.japanese,Hayashi.conference}, the quantum version of LAN has been derived in a series of papers \cite{Guta&Kahn,Guta&Jencova,Guta&Kahn2,Guta&Janssens&Kahn} to which we refer for the details of the constructions. Here we only mention the general result, and discuss in more detail the special case of pure states models which is more relevant for the present work. 

As in the classical case, we can localise $\theta$ to a neighbourhood of $\theta_{0}$ such that $\theta= \theta_{0}+u/\sqrt{n}$ with $\|u\|\leq n^{\epsilon}$ for some small 
$\epsilon>0$, and we denote $\rho_{n,u}:=\rho_{\theta_{0}+u/\sqrt{n}}^{\otimes n}$. Then there exist channels (normalised, completely positive linear maps) $T_{n}$ and $S_{n}$ between the appropriated spaces such that 
\begin{eqnarray}
&&
\lim_{n\to\infty}
\sup_{\|u\|\leq n^{\epsilon}} 
\| T_{n}(\rho_{n,u})  -  N_{u} \otimes\Phi_{u} \|_{1}
=0,
\label{eq.tn}\\
&&
\lim_{n\to\infty}
\sup_{\|u\|\leq n^{\epsilon}} 
\| \rho_{n,u}- S_{n}(  N_{u} \otimes\Phi_{u})\|_{1}=0
\label{eq.sn}
\end{eqnarray}
where $\{\Phi_{u}: u\in \mathbb{R}^{p}\} $ is a family of quantum Gaussian states of a continuos variables system, and $\{N_{u}: u\in \mathbb{R}^{p}\} $ is a family of classical Gaussian distributions. Moreover, each family has a fixed covariance matrix and the displacement is a linear transformation of the unknown parameter $u$. With this tool at hand, one can prove that the Holevo bound is attainable through the following three steps procedure: first localise 
$\theta$, then send the remaining states through the channel $T_{n}$ and then apply the optimal measurement for the limit Gaussian model.

We illustrate LAN for the simple case of a one dimensional family of pure states on $\mathbb{C}^{d}$
\begin{equation}\label{eq.unitary.family}
|\psi_{\theta}\rangle:= \exp(-i\theta J) |\psi\rangle, 
\end{equation}
where $J$ is a selfadjoint operator, which satisfies 
$\langle\psi |J| \psi\rangle=0$. The (non-unique) sld is given by
$$
\mathcal{L}_{\theta}= -2 i [ J, |\psi_\theta\rangle\langle \psi_{\theta}| ] 
$$
 and the quantum Fisher information is proportional to the variance of the `generator' $J$
\begin{equation}\label{eq.usual.fisher}
F(\theta)=    \langle \psi_{\theta}|  \mathcal{L}_{\theta}^{2} |\psi_{\theta}\rangle= 4 \langle \psi|  J^{2} |\psi\rangle:=F.
\end{equation}

In the case of pure states, the limit Gaussian model consists of a family of  coherent states so that LAN can be intuitively understood by analysing the intrinsic geometric structure of the quantum statistical model, which is encoded in the inner products between vectors corresponding to different parameters. Indeed, with the usual definition of the local states 
$|\psi_{n,u}\rangle:= |\psi_{\theta_{0}+ u/\sqrt{n}}\rangle^{\otimes n}$, a simple calculation shows that
\begin{eqnarray}
\lim_{n\to\infty} &&\langle \psi_{n,u}| \psi_{n,v}\rangle = 
\lim_{n\to\infty}  \langle \psi |\exp(i(u-v)/\sqrt{n}J) | \psi\rangle^{n}
\nonumber\\
&&=
\exp(-(u-v)^{2} F/8)
=\langle \sqrt{2F}u |\sqrt{2F}v\rangle\label{eq.weak.lan}
\end{eqnarray}
where $ |u\rangle$ is a coherent state of a one mode continuous variables (cv) system, with displacements 
$\langle u|Q|u \rangle=u$ and $\langle u|P|u\rangle=0$. This means that locally, the  `shape' of the statistical model for $n$ states, converges to that of a family of coherent states. By using the central limit, we can identify the collective observables which converge ``in distribution'' to the coordinates of the cv system as
\begin{eqnarray*}
\sqrt{\frac{2}{n F}} J(n)&:=&\sqrt{\frac{2}{n F}}
\sum_{i=1}^{n} J^{(i)} \longrightarrow
P\\
\frac{1}{\sqrt{2n F}} \mathcal{L}_{\theta_{0}}(n)&:=& \frac{1}{\sqrt{2n F}}\sum_{i=1}^{n} \mathcal{L}_{\theta_{0}}^{(i)} \longrightarrow
Q\\
\end{eqnarray*}
such that the rescaled sld $\mathcal{L}_{\theta_{0}}(n)/\sqrt{n}$ converges to the sld of the limit Gaussian model $\{|\sqrt{2F}u\rangle : u\in\mathbb{R}\}$, as expected. With a more careful analysis of the speed of convergence in \eqref{eq.weak.lan}, it can be shown that the weak LAN can be upgraded to the strong version described by \eqref{eq.tn} and \eqref{eq.sn} \cite{Guta&Bouten}. 

The goal of  this paper is to derive the weak LAN for the output state 
of a mixing Markov chain together with its classical counterpart for averages of simple measurements. The discussion around the i.i.d. models will hopefully provide the necessary intuition about the statistical meaning of LAN  in the Markov set-up and convince the reader that the quantity \eqref{eq.q.fisher.info} plays the role of asymptotic quantum Fisher information per atom. We leave the purely technical step for of deriving the strong LAN, and proving the achievability of the quantum Fisher information for \cite{Guta&Bouten}.

\section{Mixing quantum Markov chains} 
For later purposes we recall some ergodicity notions for a Markov chain with fixed unitary $U$. The reduced n-steps dynamics of the cavity is given by the CP map $\rho\mapsto T_{*}^{n}(\rho)$ where $T_*$ is the `transition operator' $T_*(\rho)= {\rm Tr}_{a} (U\, |\psi \rangle\langle\psi|\otimes \rho \, U^{\dagger})$ with the trace taken over the atom. We say that $T_*$ is {\it mixing} if it has a unique stationary state $T_{*}(\rho_{st}) = \rho_{st}$, and any other state converges to $\rho_{st}$ i.e. 
$
 \|T_*^{n}(\rho) -\rho_{st}\|_{1} \to 0.
$  
Mixing chains have a simple characterisation in terms of the eigenvalues of $T_*$, generalising the classical Perron-Frobenius Theorem.
\begin{thm}
$T_*$ is mixing if and only if it has a unique eigenvector with eigenvalue 
$\lambda_{1}=1$ and all other eigenvalues satisfy $|\lambda|<1$. 
\end{thm}
As a corollary, the convergence to equilibrium is essentially exponentially fast 
$\|T^{n}(\rho) -\rho_{st}\|_{1} \leq C_{k} n^{k} |\lambda_{2}|^{n}$ where 
$\lambda_{2}$ is the eigenvalue of $T$ with the second largest absolute value 
\cite{Terhal&DiVincenzo}. The following theorem is a discrete time analogue of the perturbation theorem 5.13 of \cite{Dav80} and its proof is given in the appendix.
\begin{thm}\label{Th.perturbation.semigroup}
Let $T(n):M(\mathbb{C}^k)\to:M(\mathbb{C}^k)$ be a sequence of linear contractions with asymptotic expansion
\begin{equation}\label{eq.t(n)}
T(n) = T_0 +\frac{1}{\sqrt{n}} T_1 + \frac{1}{n} T_2 + O(n^{-3/2}),
\end{equation}
such that $T_0$ is a mixing CP map with stationary state $\rho_{st}$. Then
${\rm}Id -T_{0}$ is invertible on the orthogonal complement of $\mathbf{1}$ with respect to the inner product $\langle A,B\rangle_{st}:= {\rm Tr}(\rho_{st} A^{\dagger} B)$. Assuming $\langle \mathbf{1}, T_1(\mathbf{1}) \rangle_{st} =0$  we have 
$$
\lim_{n \to\infty} T(n)^n [\mathbf{1}]= \exp(\lambda) \cdot \mathbf{1},
$$
where 
$
\lambda =  {\rm Tr}
\left(\rho_{st} 
(
T_2(\mathbf{1}) + 
T_1\circ ({\rm Id} - T_0)^{-1} \circ T_{1}(\mathbf{1} )
\right).
$
\end{thm}
From now on we assume that $T=T_{\theta}$ is mixing.
Since the cavity equilibrates exponentially fast we can choose the stationary state 
$\rho^{\theta}_{st}$ as initial state, without affecting the asymptotic results below.


\section{Main results}
This section contains the main results of the paper. The first subsection deals with estimators based on outcomes of separate identical measurements on the output atoms. In Theorem \ref{th.simple.measurement} we prove the asymptotic normality of the outcomes' time average, and find the explicit expression of the asymptotic Fisher information of such statistics. Similar results can be obtained for time averages of functionals depending on several outcomes, such as the correlations between subsequent atoms. In general these will provide higher Fisher information since the measurement outcomes are {\it not} independent but have exponentially decaying correlations.  It remains an open problem is to find the `full' Fisher information of the measurement stochastic process. The second subsection deals with the intrinsic statistical properties of the quantum model $\psi^{n}_{\theta}$. In Theorem \ref{th.qlan} we prove that the quantum model is asymptotically normal, and we find the explicit expression of the quantum Fisher information $F$ per atom.

\subsection{Simple measurements}

Consider a simple measurement scheme where an observable $A\in M(\mathbb{C}^{d})$ is measured on each of the outgoing atoms, and let 
${\bf A}^{(l)}$ be the random outcome of the measurement on the $l$'s atom. 
By stationarity, all expectations $\mathbb{E}_{\theta}({\bf A}^{(l)})$ are equal to 
$$
\langle A\rangle_{\theta}:= {\rm Tr}\left( |\psi\rangle\langle\psi| \otimes \rho^{\theta}_{st} \, U_{\theta}^{\dagger}(A\otimes \mathbf{1}) U_{\theta} \right)
$$ 
and by ergodicity of the measurement process \cite{Maassen&Kummerer} 
$$
\bar{\bf A}_n:= \frac{1}{n} \sum_{l=1}^{n} {\bf A}^{(l)} \to \langle A\rangle_{\theta}, \qquad  a.s.
$$
Generically the right side depends smoothly on $\theta$ and can be inverted (at least locally) so that $\theta= f(\langle A\rangle_{\theta})$, for some well behaved function $f$, providing us with the estimator $\hat{\theta}_{n} := f(\bar{\bf A}_n)$. As argued above, to analyse its asymptotic performance  
we can take $\theta=\theta_{0}+u/\sqrt{n}$ with $\theta_{0}$ fixed and $u$ an unknown `local parameter'.
\begin{thm}\label{th.simple.measurement}
Let $\theta= \theta_0+ u/\sqrt{n}$ with $\theta_0$ fixed and let 
$A\in M(\mathbb{C}^d)_{sa}$ be such that $\langle A \rangle_{\theta_0}=0$. 
Then $\sqrt{n}\bar{\bf A}_{n}$ is asymptotically normal, i.e. as $n\to \infty$
$$
\mathbb{E}_{\theta_0+u/\sqrt{n}}\left[\exp((it \sqrt{n}\bar{\bf A}_n)\right]
\to F_u(t)
$$
where 
$F_u(t) :=\exp\left(i\mu(A)u t -\frac{\sigma^{2}(A)t^2}{2} \right)$ 
is the characteristic function of the distribution $N(\mu(A)u, \sigma^2)$ with
\begin{align}
\mu(A)&:= i \langle [H, A\otimes \mathbf{1} +\mathbf{1}\otimes B  ] \rangle_{\theta_0}\\
\sigma^2(A)&:= \langle A^2\rangle_{\theta_0} + 2\langle A\otimes B\rangle_{\theta_0} \label{eq.variance.clt}\\
B&:= ({\rm Id} - T_0)^{-1}  
( \langle\psi | U_{\theta_0}^\dagger A\otimes \mathbf{1} U_{\theta_0} |\psi\rangle)\label{eq.B}
\end{align}
\end{thm}
Note that for $u=0$ we obtain a quantum extension of the Central Limit Theorem for Markov chains \cite{Meyn&Tweedie}.

Before proving the theorem we show that 
$\hat{\theta}_{n}$ is asymptotically normal and find its mean square error. 
By expanding $\hat{\theta}_{n}:= f( \bar{\bf A}_{n})$ around $\langle A\rangle_{\theta_{0}}$ and using the property $f^{\prime}(\langle A\rangle_{\theta})= (d\langle A\rangle_{\theta} /d\theta)^{-1} $ we have
\begin{align*}
\hat{\theta}_{n}& = f(\langle A\rangle_{\theta_{0}}) + 
f^{\prime} (\langle A\rangle_{\theta_{0}}) 
( \bar{\bf A}_{n} -\langle A\rangle_{\theta_{0}} ) + O_{\mathbb{P}}(n^{-1})\\
&= \theta_{0}+ \mu(A)^{-1} \bar{\bf A}_{n} + O_{\mathbb{P}}(n^{-1})
\end{align*}
so that 
$$
\lim_{n\to\infty}n\mathbb{E}\left[ ( \hat{\theta}_{n} - \theta )^{2}\right] =  
\lim_{n\to\infty}
\mathbb{E} \left[  \left( \frac{ \sqrt{n}\bar{\bf A}_{n} }{\mu(A)}  - u \right)^{2}\right]= \frac{\sigma^2(A)}{\mu(A)^2}.
$$
The limit can be seen as the inverse (classical) Fisher information per measured atom, which in principle can be minimized by varying $A$ and/or the input state $\psi$. 
\begin{proof} By using \eqref{eq.output} we can rewrite the characteristic function as 
\begin{align*}
\mathbb{E}_{\theta}\left[\exp((it \sqrt{n}\bar{\bf A}_n)\right]
& =
\left\langle \exp(it \sqrt{n}\bar{A}_{n}) \right\rangle_\theta = 
\left \langle T(n)^n [\mathbf{1}]) \right \rangle_\theta
\end{align*} 
where $T(n):M(\mathbb{C}^{k})\to M(\mathbb{C}^{k})$ is the map
$$
T(n):X\mapsto 
\mathbb{E}_{\theta_{0}}
\left.\left[ 
e^{iHu/\sqrt{n}} \left( e^{itA/\sqrt{n}}\otimes X \right)e^{-iHu/\sqrt{n}} \right| {\bf s} \right]
$$
with $\mathbb{E}_{\theta_{0}}\left[ Y|{\bf s}\right] := \langle \psi|U_{\theta_{0}}^{\dagger} Y U_{\theta_{0}}| \psi\rangle$ the `conditional expectation' onto the system. We expand $T(n)$ as in \eqref{eq.t(n)} with
\begin{align*}
T_{0}(X):=& \mathbb{E}_{\theta_{0}}\left[  \mathbf{1}\otimes X |{\bf s}\right]
\\
T_{1}(X) :=&i \mathbb{E}_{\theta_{0}}\left[ u[H,\mathbf{1}\otimes X]+t  A\otimes X|{\bf s}\right] 
\\
T_{2}(X):=& -\frac{u^{2}}{2}\mathbb{E}_{\theta_{0}}\left[ [H,[H, \mathbf{1}\otimes X]] |{\bf s}\right] 
\nonumber\\
&\mbox{}~-\frac{t^{2}}{2}\mathbb{E}_{\theta_{0}}\left[A^{2}\otimes X|{\bf s}\right] 
- ut \mathbb{E}_{\theta_{0}}\left[[H, A\otimes X]|{\bf s}\right] 
 \end{align*}
Since ${\rm Tr}(\rho_{st} T_{1}(\mathbf{1}))= \langle A\rangle_{\theta_{0}}=0$ we can apply Theorem \ref{Th.perturbation.semigroup} so we only need to compute the coefficient $\lambda$. The first  part is 
${\rm Tr}(\rho_{st}T_{2}(\mathbf{1}) )= -t^{2}\langle A^{2}\rangle_{\theta_{0}}/2 -ut\langle [H, A\otimes \mathbf{1}]\rangle_{\theta_{0}} $ and the second part is $it {\rm Tr}(\rho_{st}\cdot T_{1}(B))$, with $B$ defined 
in \eqref{eq.B}.

\end{proof}

\subsection{Quantum Fisher information and LAN}

Recall that the joint state of the system and output atoms is
\begin{eqnarray*}
\psi^{n}_{\theta}& = &U_{\theta}(n) \, \left(\psi^{\otimes n}\otimes \varphi \right):= 
U^{(1)} \cdots U^{(n)} \, \left(\psi^{\otimes n}\otimes \varphi\right)\\
&=& \exp(-i\theta H^{(1)})\cdots  \exp(-i\theta H^{(n)}) \, \left(\psi^{\otimes n}\otimes \varphi\right)
\end{eqnarray*}
where $H^{(i)}$ represents the copy of the interaction hamiltonian which acts on the system and the $i$'th atom, and is ampliated by the identity on the rest of the atoms. It is important to note that in general the commutants $[H^{(i)}, H^{(j)}]$ are nonzero for $i\neq j$ since both hamiltonians contain system operators. This means that the model 
$\psi^{n}_{\theta}$ is not a covariant one as \eqref{eq.unitary.family}, i.e. we cannot write it as
$\psi^{n}_{\theta} = \exp(-i\theta \tilde{H}(n)) \left(\psi^{\otimes n}\otimes \varphi\right)$  for some `total hamiltonian' 
$\tilde{H} (n)$. In particular, as we will see, the quantum Fisher information depends on $\theta$. 
However, we can write
$$
\frac{d\rho^{n}_{\theta}}{d\theta} =   \left[  H_{\theta}(n), \rho^{n}_{\theta} \right] 
 $$
where $\rho^{n}_{\theta}= |\psi^{n}_{\theta}\rangle\langle \psi^{n}_{\theta}| $ and  
$$
H_{\theta}(n) = \sum_{i=1}^{n} H^{(i)}_{\theta}  := \sum_{i=1}^{n} U_{\theta}^{(1)}\cdots U_{\theta}^{(i-1)} H^{(i)}  U_{\theta}^{(i-1)*} \cdots  U_{\theta}^{(1)*}  .
$$
As in \eqref{eq.usual.fisher} it follows that the quantum Fisher information $F^{(n)} (\theta)$ is
$$
F^{(n)} (\theta) = 4 \left<  H_{\theta}(n)^{2} \right>_{\theta,n} - 4 \left<  H_{\theta}(n)\right>_{\theta,n}^{2}.
$$
Here and in the next few lines the expectation $\langle\cdot \rangle_{\theta,n}$ is taken with respect to the state 
$\psi^{n}_{\theta}$, rather than the stationary state whose expectation is denoted $\langle\cdot \rangle_{\theta}$. 
We will see that in asymptotics, we can revert to the stationary state expectation.  Ignoring for the moment the second term on the right side, we write
\begin{eqnarray*}
\left<  H_{\theta}(n)^{2} \right>_{\theta,n} & = & \sum_{i=1}^{n} \left\langle \left(H^{(i)}_{\theta}\right)^{2} \right\rangle_{\theta,n}\\
&&+ 
\sum_{ 1\leq i< j \leq n}  \left \langle  \left\{H^{(i)}_{\theta} , H^{(j)}_{\theta}\right\} \right \rangle_{\theta,n}.
\end{eqnarray*}
Now, by ergodicity we have
\begin{eqnarray*}
&&
\lim_{n\to\infty} \frac{1}{n} \sum_{i=1}^{n}  \left\langle \left(H^{(i)}_{\theta}\right)^{2} \right\rangle_{\theta,n} = 
\left\langle H^{2} \right\rangle_{\theta}.
\end{eqnarray*}
Similarly, by using the Markov property, it can be shown that 
$$
\lim_{n\to\infty} \frac{1}{n}\sum_{ 1\leq i< j \leq n}  \left \langle  \left\{H^{(i)}_{\theta} , H^{(j)}_{\theta}\right\} \right \rangle_{\theta,n}= \sum_{i=0}^{\infty} \left\langle   \left\{ H  ,T_{0}^{i} ( K) \right\} \right\rangle_{\theta}.
$$
where  $K:=\langle \psi | H |\psi \rangle\in M(\mathbb{C}^{k})$. 
Assuming that $\langle H\rangle_{\theta}=0$, the bias term $\langle H_{\theta}(n) \rangle_{\theta,n}^{2}$ is sub-linear and from the above limits we obtain 
\begin{equation}\label{eq.limit.q.Fisher}
\lim_{n\to\infty} \frac{1}{n} F^{(n)}(\theta_{0}) = F(\theta_{0})
\end{equation}
with $F(\theta_{0})$ the asymptotic quantum Fisher information per atom, given by \eqref{eq.q.fisher.info}. The next theorem strengthens this conclusion, by showing that the quantum model $\psi^{n}_{\theta}$ converges (locally around any 
$\theta_{0}\neq 0$) to a coherent state model with quantum Fisher information $F(\theta_{0})$. 
 
\begin{thm}\label{th.qlan}
Suppose that $\langle H\rangle_{\theta_{0}}=0$. Then the family of (local) output states 
$
\{ \psi_{\theta_0+u/\sqrt{n}}^n :u\in \mathbb{R}\}$
converges to a family of coherent states $\{|\sqrt{F/2}u\rangle : u\in \mathbb{R} \}$ with 1-d
 displacement $\langle \sqrt{F/2}u| Q| \sqrt{F/2}u\rangle=\sqrt{F/2}u $. More precisely, as $n\to \infty$
\begin{align}\label{eq.model.convergence}
\langle \psi_{\theta_0+u/\sqrt{n}}^n | \psi_{\theta_0+v/\sqrt{n}}^n  \rangle\to &
e^{i a (u^{2}-v^{2})}\langle \sqrt{F/2}u | \sqrt{F/2}v\rangle\nonumber\\
&= e^{i a (u^{2}-v^{2})} e^{-F(u-v)^2/8}
\end{align}
where $a\in \mathbb{R}$ is a constant and $F = F(\theta_{0})$ is the asymptotic quantum Fisher information (per atom)
\begin{align}\label{eq.q.fisher.info}
F(\theta_{0})&:= 4\left[\langle H^2 \rangle_{\theta_0} +\left\langle \{ H, ({\rm Id}-T_0)^{-1}(K ) \} \right.\rangle_{\theta_0}\right],
\end{align}
with $K:= \mathbb{E}_{\theta_{0}} [H|{\bf s}]=\langle \psi | H |\psi \rangle$.
\end{thm}

Note that $e^{ia(u^{2}-v^{2})}$ is an irrelevant phase factor which can be absorbed in the definition of the limit states. 
The claim that $F$ is the asymptotic quantum Fisher information (per atom) of the output follows 
from the convergence \eqref{eq.model.convergence}  to the one dimensional coherent state model $\{ |\sqrt{F/2}u\rangle :u\in \mathbb{R}\}$ and the fact that the latter has quantum Fisher information $F$, cf. \cite{Helstrom,Braunstein&Caves}, and agrees with the limit \eqref{eq.limit.q.Fisher}. Note also that similarly to \eqref{eq.variance.clt}, 
the left side of \eqref{eq.q.fisher.info} can be identified with the asymptotic variance appearing in the CLT for the operator $2H$. This agrees with the formula of the quantum Fisher information for unitary families of pure 
states, as variance of the driving hamiltonian as shown in section \ref{sec.2}. 
By extending the weak convergence to strong convergence as described in section \ref{sec.2} and applying similar techniques as in the i.i.d. case \cite{Guta&Janssens&Kahn} it can be shown that
there exists a two step adaptive measurement which asymptotically achieves the smallest possible variance 
\begin{equation}\label{eq.variance.optimal}
n\mathbb{E}\left[( \hat{\theta}_{n}-\theta)^{2}\right] \to F^{-1},\quad \hat{\theta}_n:= \theta_0+ \hat{u}_{n}/\sqrt{n}.
\end{equation}
Since for coherent models the optimal measurement is that of the canonical variable which is conjugate to the `driving' one, we conjecture that quantum Fisher information is achieved by a measuring a collective observable which is conjugate to the hamiltonian $H$ in a more general Markov CLT than that of Theorem \ref{th.simple.measurement}. 

{\it Proof of Theorem 4.} By \eqref{eq.output} 
we can rewrite the inner product as
$$
\langle \psi_{\theta_0+u/\sqrt{n}}^n | \psi_{\theta_0+v/\sqrt{n}}^n  \rangle = 
{\rm Tr} (\rho T(n)^{n}[\mathbf{1}])
$$
where 
$$
T(n):X\mapsto \mathbb{E}_{\theta_{0}} \left. \left[ e^{iuH/\sqrt{n}}( \mathbf{1}\otimes X) e^{-ivH/\sqrt{n}}\right|\mathbf{s} \right]. 
$$
Its expansion is  
\begin{align*}
T_{0}(X):=& \mathbb{E}_{\theta_{0}}\left[  \mathbf{1}\otimes X |{\bf s}\right]
\\
T_{1}(X) :=&i \mathbb{E}_{\theta_{0}}\left[ u H(\mathbf{1}\otimes X ) -v(\mathbf{1}\otimes X )H |{\bf s}\right] 
\\
T_{2}(X):=& 
-\frac{1}{2}
\mathbb{E}_{\theta_{0}}\left[
u^{2} H^{2}(\mathbf{1}\otimes X ) + 
v^{2}(\mathbf{1}\otimes X) H^{2}|{\bf s}\right] 
\nonumber\\
&\mbox{}~+
\mathbb{E}_{\theta_{0}}\left[uv H(\mathbf{1}\otimes X )H |{\bf s}\right].
\end{align*}
and the condition of Theorem 2 
holds: ${\rm Tr}(\rho_{st} T_{1}(\mathbf{1})) = \langle H\rangle_{\theta_{0}}=0$.  Since $T_{1}(\mathbf{1})=i(u-v)K$, the contribution from ${\rm Tr}(\rho_{st} T_{1}\circ ({\rm Id}- T_{0})^{-1}\circ T_{1}(\mathbf{1})) $ can be written as
\begin{align*}
&-(u-v)^{2}\, {\bf Re}\langle H  ({\rm Id}- T_{0})^{-1}(K) \rangle_{\theta_{0}}  \\
&-i(u^{2}-v^{2})\, {\bf Im}\langle H  ({\rm Id}- T_{0})^{-1}(K) \rangle_{\theta_{0}} 
\end{align*}
Moreover, ${\rm Tr}(\rho_{st} T_{2}(\mathbf{1}))=-(u-v)^{2} \langle H^{2} \rangle_{\theta_{0}}/2$, so adding the two contributions we obtain the desired result.

\qed

%
%
%
%
%

\section{Example} 
In this section we illustrate the main results on the classical and quantum Fisher informations, for a simple discrete time $XY$-interaction model. We also analyse the behaviour of the quantum Fisher information in the vicinity of the point $\theta=0$ at which the chain is not ergodic, and find that it is quadratic rather than linear in $n$ and is concentrated in the system rather than the output.

\begin{center}
\begin{figure}
\includegraphics[width=6.cm]{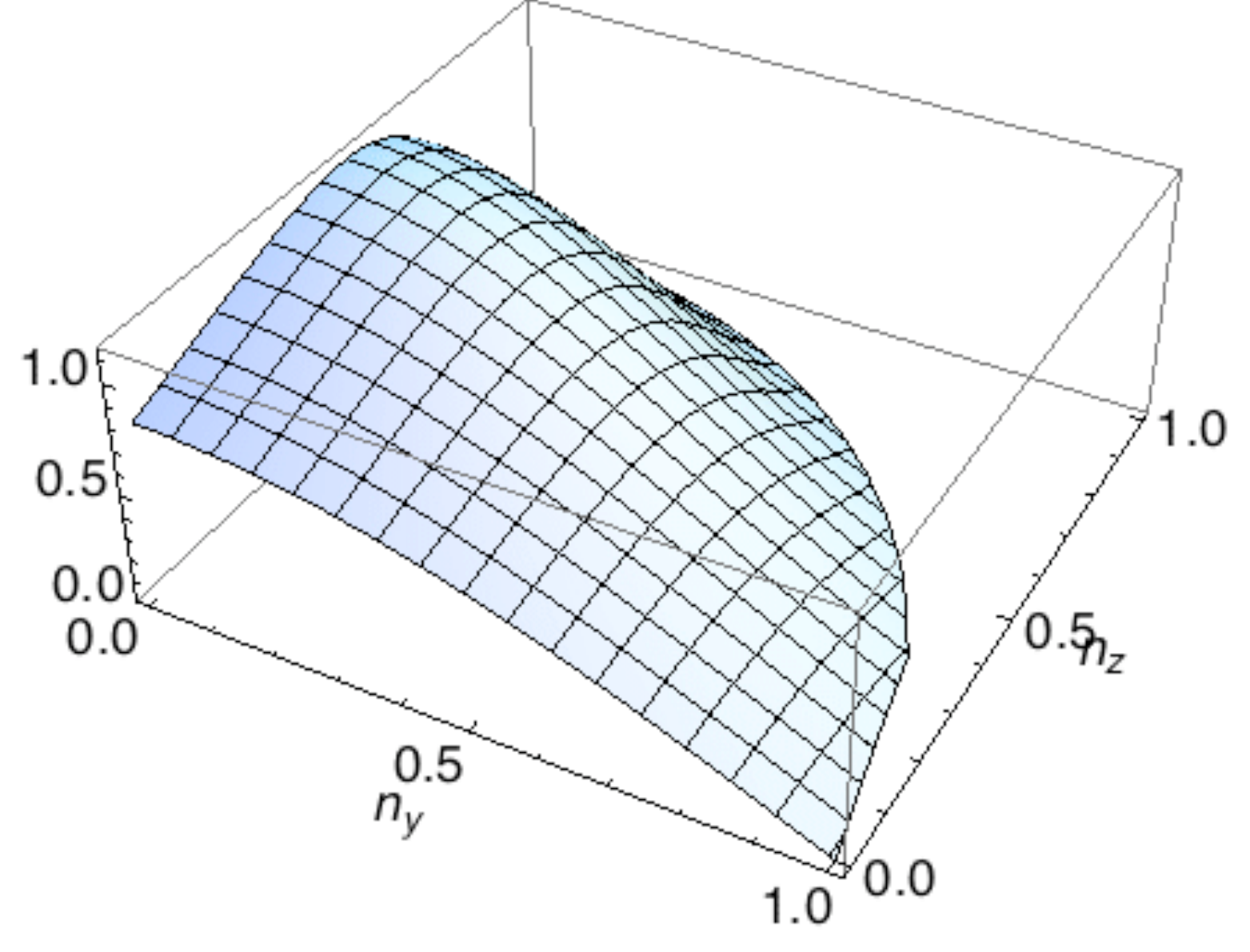}
\caption{Classical Fisher information for spin measurements as function of spin components $n_{y},n_{z}$ }\label{fig.fisher}
\end{figure}
\end{center}

Let $U_{\theta}= \exp(-i\theta H)$ with $H$ the 2-spin `creation-annihilation' hamiltonian 
$H=i(\sigma_{+}\otimes \sigma_{-} - \sigma_{-}\otimes \sigma_{+}) $ where $\sigma_{\pm}$ are the raising and lowering operators on $\mathbb{C}^{2}$, and let  
$|\psi\rangle\ := a | 0\rangle + be^{if} |1\rangle$ be the input state, with $a^{2}+b^{2}=1$. 
The transition matrix $T_{0}$ has eigenvalues 
$\{1,c, (c( c+1) \pm \sqrt{c^{2}(1-c)^{2} - 16a^2b^{2}c(1-c^{2})} ) /2\}$, and is mixing if and 
only if $c:=\cos\theta_{0}\neq 1$. The quantum Fisher information \eqref{eq.q.fisher.info} is
\begin{equation}\label{eq.Fisher.example}
F:= \frac{ 16 a^{4} b^{4} }{ (1-c) (1-c +4a^{2} b^{2}c )}
\end{equation}
which is independent of the phase $f$. This can be compared with the value of the classical Fisher information $\mu(\sigma_{\vec{n}})^2/\sigma^2(\sigma_{\vec{n}})$ obtained by measuring  the spin in the direction $\vec{n}$ for each outgoing atom, cf. Theorem \ref{th.simple.measurement}. For $c=0.5, f=0,b=0.8$ the quantum Fisher information is $F=5.03$ while that of the spin measurement varies from $0$ to $1.15$ (see Figure \ref{fig.fisher}).
An interesting, and perhaps surprising feature of the quantum Fisher information \eqref{eq.Fisher.example} is that it diverges at vanishing coupling constant, due to the factor $1-\cos(\theta)\approx \theta^{2}/2$ in the denominator. This singularity arises from the second term on the right side 
of \eqref{eq.q.fisher.info} and stems from the fact that  the Markov chain is 
{\it not mixing} at $\theta=0$. To get an intuition for this phenomenon let us compute (according to the standard methodology \cite{Helstrom,Holevo,Braunstein&Caves})  the quantum Fisher information at 
$\theta=0$ for the family
$$
\psi_{\theta}^{n}:= U_{\theta}(n)( \psi^{\otimes n}\otimes\varphi),
$$
where the input state is chosen to be $|\psi\rangle=(|0 \rangle +|1\rangle)/\sqrt{2}$ and the initial state of the `cavity' is $|\varphi\rangle =|0\rangle$. 
Since
$$
- i H(n) := \left. \frac{dU_{\theta}(n)}{d\theta}\right|_{\theta=0} = \sum_{i=1}^{n}
\left(\sigma_{+}\otimes \sigma^{(i)}_{-} - \sigma_{-}\otimes \sigma^{(i)}_{+}\right)
$$  
one can easily verify that 
$$
\langle \psi_{\theta=0}^{n} | H(n) | \psi_{\theta=0}^{n}\rangle=0,
$$
and hence the quantum Fisher information at $\theta=0$ is 
$$
F^{(n)} = 
4\langle  \psi_{\theta=0}^{n} | H(n)^{2}| \psi_{\theta=0}^{n}\rangle = 
4\| H(n)  | \psi_{\theta=0}^{n}\rangle \|^{2}.
$$
Since $\sigma_{-} |0\rangle=0$ the latter reduces to computing the squared norm of the vector
$$
\left( \sum_{i=1}^{n} \sigma_{-}^{i} \right) 
\left( \frac{|0\rangle + |1\rangle }{\sqrt{2}}\right)^{\otimes n}= 
\sum_{i_{1},\dots , i_{n}} 
c(i_{1},\dots i_{n}) | i_{1},\dots i_{n}\rangle.
$$
where $i_{j}\in \{0,1\}$ are the basis indices. Since a basis vector containing 
$p$ indices equal to $0$, can be obtained in $p$ different ways by applying the lowering operator to an input tensor, the coefficients are 
$$
c(i_{1},\dots, i_{n}) = (n-\sum_{j=1}^{n} i_{j})/2^{n/2}, 
$$
and the Fisher information is 
$$
F^{(n)}= 2^{-n}\sum_{p=0}^{n} p^{2} {n \choose p}= n(n+1).
$$
Thus, in contrast to the case of independent systems, the Fisher information scales {\it quadratically} rather than linearly with the number of systems, hence the divergence of the quantum Fisher information {\it per atom} which represents the asymptotic value of $F^{(n)}/n$. Note that similar quadratic scaling of the quantum Fisher information is encountered in phase estimation \cite{Giovanetti} and more generally in optimal estimation of unitary channels \cite{Kahn_sud}, with the difference that it holds for any parameter, rather than at a single point. 

We take a closer look at the the states $\psi^{n}_{\theta}$ by scaling the parameter as $\theta = u/n$ as suggested by the Fisher information. This means that we know $\theta$ with an accuracy of $n^{-1}$ (rather than the usual $n^{-1/2}$) and we would like to find if the the estimation of $u$ `stabilises' in the asymptotic regime. By using the same technique as in Theorem \ref{th.qlan} we have
\begin{equation}\label{eq.innerproducts}
\langle \psi^{n}_{u/n}|  \psi^{n}_{v/n}\rangle = \langle 0 |T(n)^{n}[\mathbf{1}]|0 \rangle
\end{equation} 
where $T(n): M(\mathbb{C}^{k})\to M(\mathbb{C}^{k})$ is the map
$$
T(n):
X\mapsto 
\langle \psi|  e^{iuH/n}( \mathbf{1}\otimes X) e^{-ivH/n}|\psi \rangle. 
$$
We expand $T(n)$ as 
\begin{equation*}
T(n)= Id + \frac{T_{0}}{n} + o(n^{-2}),
\end{equation*}
where 
\begin{equation*}
T_{0}: X\mapsto i ( u K  X - v X K) , 
\quad K:=\langle\psi| H|\psi\rangle.
\end{equation*}
Plugging into \eqref{eq.innerproducts} we obtain the limit 
\begin{eqnarray}
\lim_{n\to\infty}
\langle \psi^{n}_{u/n}|  \psi^{n}_{v/n}\rangle &=& 
\langle 0 | \exp(T_{0})[\mathbf{1}]|0\rangle \nonumber \\
&=& \langle 0 | \exp(i  (u-v)K) |0\rangle.\label{eq.unitary1}
\end{eqnarray}
This agrees with the quadratic scaling of the quantum Fisher information at 
$\theta=0$ and shows that the quantum statistical model has a limit provided that the right scaling of parameters is used.

Let us now look at the $n$-steps reduced dynamics of the system for the same scaling $\theta=u/n$ with initial state $|0\rangle$. 
After a similar computation (with $u=v$) we obtain 
\begin{eqnarray}
\lim_{n\to\infty} \rho(n) &=& \lim_{n\to\infty}T_{*}(n)^{n}(|0 \rangle\langle 0|) 
\nonumber\\
&=&
\exp(-iu K)|0 \rangle\langle 0|\exp(iu K),\label{eq.unitary}
\end{eqnarray}
which means that for large $n$ the system is effectively {\it unitarily driven} by the  `hamiltonian' $K$, even though we started with an open system dynamics! This effect is interesting in itself and is reminiscent of the quantum Zeno effect. From \eqref{eq.unitary1} and \eqref{eq.unitary} we can conclude that asymptotically, the system and the output have pure states, and moreover the input passes undisturbed into the output
$$
|\psi^{n}_{u/n} \rangle \approx |\psi\rangle^{\otimes n}\otimes \exp(iu K )|0\rangle.
$$ 
In conclusion, unlike the ergodic case where the output contains information about the parameter, it is the system's state which carries all the information, and successive time steps amount to a simple unitary rotation. 
In particular, this explains the quadratic scaling of the quantum Fisher information $F^{(n)}$. In conclusion, the non-ergodic set-up exhibits interesting statistical features and should be analysed on its own in more detail. In particular, for practical applications it is important to see whether the quadratically scaled Fisher information is achievable asymptotically.

\section{Conclusions and outlook} 
Quantum system identification is an area of significant practical relevance with interesting statistical problems going beyond the state estimation framework. We showed that in the Markovian 
set-up this problem is very tractable thanks to the asymptotic normality satisfied 
by the output state {\it and} the time average of the simple measurement process. This may come as a surprise considering that the output is correlated, but is in perfect agreement with the classical theory of Markov chains where similar results hold \cite{Hopfner}.  The theorems can be extended to strong LAN with multiple parameters, continuos time dynamics, and measurements on several atoms \cite{Guta&Bouten}. However, as in process tomography, full identification of the unitary requires the preparation of different input states. The optimisation of these states, and  the case of non-mixing chains are interesting open problems. Another open problem is to find the classical Fisher information of the simple measurement {\it process}, rather than that of time averaged functionals. Since Markov chains are closely related to matrix product states \cite{Fannes&Nachtergale&Werner,Schon&Verstrate&Wolf}, our results are also relevant for estimating matrix product states \cite{Cramer}. 

When the chain is not-ergodic, the Fisher information may exhibit a qualitatively different behaviour, such as quadratic rather than linear scaling with the number of output systems, which bears some similarity with  that encountered in phase estimation \cite{Giovanetti}. Understanding this behaviour in a more general scenario, and the possible applications in precision metrology are  topics for future research.


\begin{acknowledgments} 
This work is supported by the EPSRC Fellowship EP/E052290/1. The author thanks Luc Bouten for many discussion and his help in preparing the paper.
\end{acknowledgments}

\appendix
\section*{Appendix: Proof of Theorem \ref{Th.perturbation.semigroup}}

Since $T_{0}$ is a mixing CP-map, the identity is the unique eigenvector with 
eigenvalue $1$ and all other eigenvalues have absolute values strictly smaller than $1$. For $n$ large enough $T(n)$ has the same spectral gap property and we denote by 
$(\lambda(n), x(n))$  its largest eigenvalue and the corresponding eigenvector such that
$
\lambda(n)\to1$ and $ x(n)\to\mathbf{1}.$

Since $T(n)$ is a contraction  
$$
\|T(n)^{n}(\mathbf{1}) - T(n)^{n}( x(n))\| \leq \|T(n)^{n}\|  \, \| \mathbf{1}- x(n)\| 
\overset{n\to\infty}{\longrightarrow} 0.
$$
On the other hand since
$T(n)^{n}( x(n)) = \lambda(n)^{n} x(n)$, we have
$
T(n)^{n}(\mathbf{1}) \to  \exp(\lambda) \mathbf{1},
$ 
provided that the following limit exists
\begin{equation*}\label{eq.exponential.limit}
\lim_{n\to\infty}  \lambda(n)^{n} := \exp(\lambda).
\end{equation*}

We prove that this is the case by using  the Taylor expasions
\begin{eqnarray*}
\lambda(n) &=& \lambda_{0} + \frac{1}{\sqrt{n}} \lambda_{1} +  
\frac{1}{n} \lambda_{2} + O(n^{-3/2}), \\
x(n)&=& x_{0}+ \frac{1}{\sqrt{n}} x_{1} + \frac{1}{n} x_{2} +  O(n^{-3/2}).
\end{eqnarray*}
Then we can solve the eigenvalue problem in successive orders of approximation
\begin{eqnarray}
T_{0} (x_{0}) &=& \lambda_{0} x_{0},\nonumber\\
T_{0}(x_{1})+ T_{1}(x_{0}) &=& \lambda_{0}x_{1}+ \lambda_{1}x_{0},\nonumber\\
T_{0}(x_{2})+ T_{1}(x_{1})+ T_{2}(x_{0}) &=&
\lambda_{0}x_{2}+ \lambda_{1}x_{1}+\lambda_{2}x_{0} ~~ (*)\nonumber.
\end{eqnarray}
From the first equation we have $\lambda_{0}=1, x_{0}=\mathbf{1}$. Inserting into the 
second equation we get
$$
(T_{0}- {\rm Id}) (x_{1}) = -(T_{1} - \lambda_{1}{\rm Id})(\mathbf{1}),
$$
and by taking inner product with $\mathbf{1}$ we obtain 
$\lambda_{1}= \langle \mathbf{1}, T_{1}(\mathbf{1})\rangle_{st}=0$, by using the assumption. Hence
$$
x_{1} = ({\rm Id} -T_{0})^{-1} \circ T_{1}(\mathbf{1}) + c\mathbf{1}
$$
where $({\rm Id} -T_{0})^{-1}$ denotes the inverse of the restriction of ${\rm Id} -T_{0}$ to the orthogonal complement of $\mathbf{1}$, and $c$ is a constant.

Similarly, from the third equation in $(*)$
$$
(T_{0} -{\rm Id})(x_{2}) =-T_{1}(x_{1}) + (\lambda_{2}{\rm Id} -T_{2})(\mathbf{1}).
$$
which implies
\begin{align*}
\lambda_{2} &=
\langle \mathbf{1}, T_{2}(\mathbf{1})\rangle_{st} + 
\langle \mathbf{1}, T_{1}(x_{1})\rangle_{st}\\
&=\langle \mathbf{1}, T_{2}(\mathbf{1})+ T_{1}\circ({\rm Id}-T_{0})^{-1}\circ T_{1}(\mathbf{1}) \rangle_{st}
\end{align*}
Finally, 
\begin{align*}
\lim_{n\to\infty}\lambda(n)^{n} &=
\lim_{n\to\infty}\left(\lambda_{0} + \frac{\lambda_{1}}{\sqrt{n}}+\frac{\lambda_{2}}{n}+O(n^{-3/2})\right)^{n}\\
&= \lim_{n\to\infty}\left(1 +\frac{\lambda_{2}}{n}\right)^{n} =\exp(\lambda_{2}).
\end{align*}

\qed


\end{document}